\documentstyle[psfig]{article}
\begin{document}

\title{\normalsize \bf DIFFUSION DUE TO THE BEAM-BEAM INTERACTION AND 
FLUCTUATING FIELDS IN HADRON COLLIDERS} 
\author{Tanaji Sen \\DESY, 85 Notkestrasse, 22607 Hamburg, Germany\\
\and
James A. Ellison\\University of New Mexico, Albuquerque, NM 87131\\ \mbox{} \\
{\small Random fluctuations in the tune, beam offsets and beam size in the 
presence of the beam-beam interaction are shown} \\
{\small to lead to significant particle diffusion and
emittance growth in hadron colliders. We find that far from resonances high }\\
{\small  frequency noise causes the most diffusion while near resonances 
low frequency noise is responsible for the large emittance }\\
{\small  growth observed. Comparison of different fluctuations
shows that offset fluctuations between the beams causes the largest }\\
{\small diffusion for particles in the beam core.}
}
\date{}

\twocolumn

\maketitle

Emittance growth due to the non-linear beam-beam interaction is a 
major concern at all hadron colliders. 
The tunes (i.e. the rotation numbers of the transverse 
betatron oscillations) are always chosen to avoid low order resonances and
the overlap of high order resonances does not create chaotic regions 
large enough to lead to significant amplitude growth.
The dynamics are qualitatively different however when the time-dependence,
both deterministic and stochastic, of parameters is included. The effects of 
deterministic tune modulation are well studied and 
removing modulation lines from the betatron 
spectrum reduces particle loss from the tails of the beam
\cite{Bru}. Random fluctuations in the tune, closed orbit and beam size
are also present in accelerators. Qualitative arguments 
\cite{Month} and numerical simulations \cite{Brink} have shown that
tune fluctuations lead to emittance growth especially for
tunes close to a resonance . Our aim is to provide a  quantitative 
theory to explain these observations.

We consider collisions of a proton beam with an opposing beam 
composed of either leptons as at HERA or hadrons as at the Fermilab
Tevatron and the proposed LHC at CERN. For one dimensional motion, the 
beam-beam potential seen by a proton, assuming a Gaussian charge distribution
of the opposing beam, is given by $U(x) = C\int_0^{\infty}dq\;
       [ 1 - e^{-x^2/(2\sigma_{op}^2+q)}] /(2\sigma_{op}^2+q)$.
The constant is $C=N_{b,op}r_p/\gamma_p$ where  $N_{b, op}$ is the number of
 particles per bunch in the opposing beam,
$r_p$ is the classical radius of the proton, $\gamma_p$ is the relativistic
kinematic factor for the protons, and $\sigma_{op}$ is the sigma of the size
of the opposing beam. Transforming to action-angle coordinates 
$(J,\psi)$ via
$x = \sqrt{2J\beta^*}\cos\psi$, $x'=-\sqrt{2J/\beta^*}\sin\psi$ where
$\beta^*$ is the $\beta$ function at the interaction point, we 
obtain the Fourier expansion of the potential 
\begin{equation}
U(x) = C[F_0(a) + 
             2 \sum_{k=1}^{\infty} F_k(a) \cos 2k\psi] \label{poten}
\end{equation}
where $a = \beta^* J/(2\sigma_{op}^2)$ is a dimensionless amplitude.
The Fourier amplitudes are 
\begin{eqnarray}
F_0 & = & \int_0^{a} \frac{1 - e^{-w}I_0(w)}{w}  dw \; , \\
F_k & = & (-1)^{k+1} \int_0^{a} \frac{e^{-w}I_k(w)}{w} dw \\
 & = & \!\! (-1)^{k+1}\frac{a^k}{2^k k k!}
   {}_2F_2[\frac{1}{2}+k,k;2k+1,k+1;-2a] \nonumber 
\end{eqnarray}
where the $I_k$ are modified Bessel functions and ${}_2F_2$ is a generalized 
hypergeometric function. Including the linear
motion and the beam-beam interaction, the Hamiltonian is 
$H=\nu^0 J + U(J,\psi)\delta_p(\theta)$ where $\delta_p(\theta)$ is the 
periodic delta function with  period $2\pi/N_{IP}$, $N_{IP}$ being the number
of interaction points and $\theta$, the "time" variable, advances by 
$2\pi$ per turn. Integrating the equations of motion over one turn leads to
the one-turn beam-beam map:
$\Delta\psi  =  2\pi\nu^0 + \partial U/\partial J $ and 
$\Delta J  =  - \partial U/\partial \psi $.

First we consider the diffusion in amplitude due to random fluctuations in the
tune. In the extreme case when the tune is completely random on the interval
[0, 1], the beam-beam kicks
occur at uncorrelated phases and the emittance grows as in a
random walk process. Usually the random contribution to the tune is quite
small, of the order of 0.001 at the most, but this is sufficient to affect the
long time dynamics. The
sources of tune fluctuation include power supply noise in quadrupoles, closed
orbit fluctuations through the non-linear magnets and mechanical vibrations of 
the non-linear magnets. In addition, both intra-beam scattering due to the
Coulomb force and RF noise lead to fluctuating particle momenta.
This in turn leads to a tune fluctuation via the machine chromaticity.
We model the tune fluctuation by an additional term 
$\Delta\psi_{ r}$ in the total phase. Assuming that the random contribution is
small, we can write the change in action 
at turn $m$ due to this fluctuating phase alone as
 $\Delta J_{ r}(m) = 
 [d\Delta J(m)/d\psi]\; \Delta\psi_{ r}(m) + O(\Delta\psi_{r}^2) $. 
The unperturbed total phase at turn $m$ and action $J$ is 
$\psi(m) = 2m\pi\nu(J)+\psi_0$, 
where $\psi_0$ is the initial phase. When the tune is far from a resonance,
the linear action $J$ is conserved after averaging. This allows
us to assume that $\Delta J_{ r}(m) \ll J(0)$ so that in the sum over turns
we can replace $J(m)$ by 
$J(0)$. We assume that the random process is stationary so that the random 
phase correlation function is of the form
 $\langle \Delta\psi_{r}(l) \Delta\psi_{r}(n+l)\rangle 
= 4\pi^2 \Delta\nu_r^2 K_{\nu}(n)$, where the average is over many realizations
of the noise process, $\Delta\nu_r$ is the amplitude of the tune fluctuations 
and $K_{\nu}(-n) = K_{\nu}(n)$. The diffusion coefficient defined as
$D_{\nu}(J) \equiv \lim_{N\rightarrow\infty} \langle [J(N) - J(0)]^2 \rangle/N$
is found, by extracting the dominant terms, to be
\begin{equation}
D_{\nu}(J) \! = \! 128 (\pi C \Delta\nu_r)^2 \sum_{k=1}^{\infty}k^4 
F_k^2 \!\!\sum_{n=-\infty}^{\infty}\!\! K_{\nu} \cos 4\pi k\nu n\; .
\end{equation}
We observe that only the tune noise at even harmonics of the betatron tune 
leads to a diffusion in the action. {\em Hence far from resonances, high
frequency components of the tune fluctuation spectrum cause diffusion.} 
A natural choice to model the tune fluctuations is the Ornstein-Uhlenbeck (OU)
process because it is the only Gaussian stationary Markov process
and the spectral density  $S(\omega)$ (related to $K_{\nu}$ by the cosine 
transform) decays as $\omega^{-2}$ which is in reasonable agreement with 
measured noise densities. For the discrete time OU process with correlation
time $\tau_c$, the correlation function is 
$K_{\nu}(n)=(1-1/\tau_c)^n/[1-1/(2\tau_c)]$.
The spectral density drops to roughly half its maximum value at a 
frequency $f_{1/2}= f_{rev}/(2\pi\tau_c)$. The revolution frequency $f_{rev}$ 
is 47.3kHz for HERA. Substituting this form for $K_{\nu}$ leads to
\begin{equation}
D_{\nu}(J) \!
 = \! 128 \frac{(\pi C \Delta\nu_r)^2}{1-1/(2\tau_c)}  \sum_{k=1}^{\infty}
     \frac{k^4 F_k^2{\rm sinh}\Theta}
     {{\rm cosh}\Theta - \cos 4\pi k\nu}\; .
  \label{Dtune}
\end{equation}
where $\Theta = - \ln(1 - 1/\tau_c)$.
The main amplitude dependence of $D_{\nu}(J)$ is contained in the
Fourier coefficients. The dominant coefficient $F_1$ has the 
expansion $F_1 = \sum_{j=1}^{\infty}(-1)^{j+1} \\
(a/2)^j \{(2j)!/[(j+1)! (j!)^2]\} $.
At small amplitudes therefore, $D_{\nu}(J) \sim J^2$. At large amplitudes the
beam-beam force vanishes and  all Fourier coefficients  $F_k$ go to constant 
values . Hence the
diffusion coefficient $D_{\nu}(J)$ increases
monotonically in the core of the beam and levels off in the tails of the beam.
Measurements of beam loss rates at HERA \cite{Seidel}   are 
consistent with a constant diffusion in the tails of the beam.
The mean escape time to an action $J_b$, calculated as $T_{esc}(J)=
\int_J^{J_b} J \; dJ/D_{\nu}(J)$, goes to infinity logarithmically at small
amplitudes as $T_{esc}(J) \sim \ln (J_b/J)$, while at large amplitudes
it goes to zero as $T_{esc}(J) \sim (J_b^2 - J^2)$.

We have compared the above analysis with a numerical calculation.
An initial distribution of 1000 particles is placed at 100 different
amplitudes with 10 particles at each amplitude distributed uniformly in phase. 
The particles are tracked for $10^7$ turns or more (the number increasing with
the noise correlation time) using the 
beam-beam map with the random contribution to the tune at each turn
determined by an OU process. The diffusion coefficients at each amplitude are
averaged over the ten phases and over ten noise realizations. The simulations
were done for three correlation times: $\tau_c=1.1, 10, 100$. 
\begin{figure}
\centerline{\psfig{figure=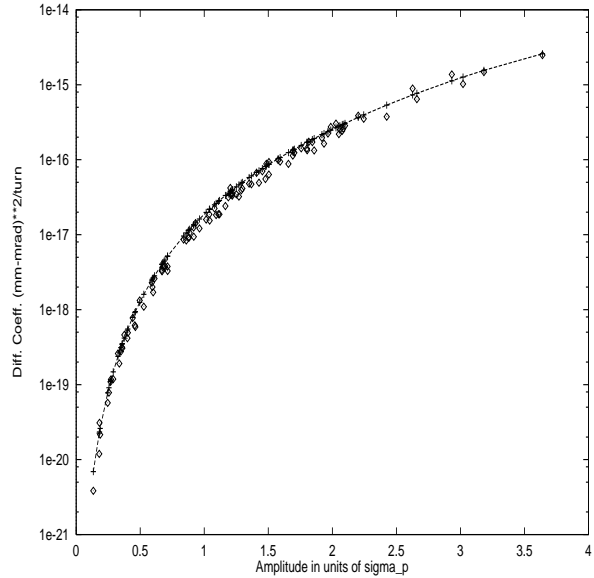,height=3.in,width=3.15in,angle=-90}}
\caption{The diffusion coefficient $D_{\nu}(J)$ calculated theoretically from
Eq. (5) (dashed line) compared with the values obtained from the simulation. 
Parameter values: $\nu^0$=0.291, $\Delta\nu_r=10^{-4}$, $\tau_c=10$. }
\end{figure}
Figure 1 shows a comparison of the diffusion
coefficient obtained from Equation (\ref{Dtune}) with that
calculated  numerically at $\tau_c=10$. The level of agreement is
nearly the same at other correlation times. Both the analysis and the 
numerical results show that 
$D_{\nu}(J) \sim \tau_c^{-1}$ for large $\tau_c$.
High frequency noise  (above 1 kHz say)
is strongly attenuated within the interior of the beam pipes
by the impedance of the magnets and other devices  in
a storage ring so realistic values of the noise correlation time for HERA are
$\tau_c \ge 10$.

The diffusion in amplitude causes emittance growth 
over the period of stored beam - typically 24 to 30 hours for
the proton beam at HERA. The time evolution of the
density and the emittance can be followed by solving the Fokker-Planck 
equation. Assuming that 
the diffusion in action is a Markov process and the drift coefficient is 
half the derivative of the diffusion coefficient (as is usual for a 
Hamiltonian system) \cite{LiLi}, the Fokker-Planck equation for the density is
\begin{equation}
\frac{\partial \rho}{\partial t} = \frac{1}{2} 
\frac{\partial}{\partial J}\left(D(J)
 \frac{\partial \rho}{\partial J}\right) \; .
\end{equation}
We integrate this one dimensional Fokker-Planck equation by the method
of lines \cite{ShihEll}. An absorbing boundary is placed at an
action $J_b$ corresponding to the position of the beam-pipe. The 
density at the origin does not change since the diffusion coefficient and its 
derivative vanish there. The evolution of the average action is then found from
$\langle J(t) \rangle = \int_0^{J_b} J \rho(J,t) dJ / \int_0^{J_b} \rho(J,t)
 dJ$.
\begin{figure}
\setlength{\unitlength}{1mm}
\mbox{
\begin{picture}(85,80)(-5,-5)
\put(-3,-4){\psfig{file=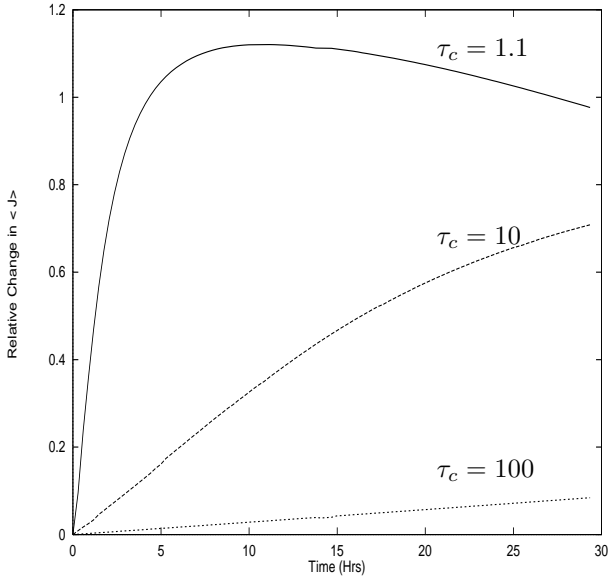,height=3.in,width=3.25in,angle=-90}}
\put(55,65){$\tau_c=1.1$}
\put(55,40){$\tau_c=10$}
\put(55,09){$\tau_c=100$}
\end{picture}
}
\caption{Relative growth in the average action $\langle J\rangle$ of a proton 
beam over a storage time of 30 hours due to tune fluctuations
at three noise correlation times.
Parameter values: $\nu^0$=0.291, $\Delta\nu_r=10^{-4}$. }
\end{figure}
Figure 2 shows the evolution of the average action over nearly
30 hours for three correlation times. For $\tau_c=1.1$, $\langle J\rangle$
grows the most rapidly as expected, then 
decreases as particles are lost at the boundary. Diffusion
is slower for the other two correlation times  and particles are not lost at
the boundary so $\langle J\rangle$ grows almost linearly with time. These 
calculations show that even in this one dimensional model,
tune fluctuation can cause the emittance to grow by 10 to 70\% over the 
storage time in a proton machine.

Emittance growth due to fluctuations is significantly enhanced 
near a resonance. 
As the tunes approach a resonance, the resonance
islands increase in width and the resonant amplitudes move further out.
Finally at the resonance tunes the fixed points have moved off to infinity 
because the beam-beam tune shift is largest at the origin and vanishes
at infinity. The phase portraits are strongly tune dependent.
For example, with $\nu=0.25$, the phase space portraits are diamond shaped 
close to the origin and at large amplitudes they are four-armed stars
with long arms along the four axes. 
A level curve in the phase portrait can be labelled by an energy to be defined 
below. Particle motion with external fluctuations has two aspects: motion
on a level curve and diffusive motion between level curves.
Near resonance a particle initially close to the origin 
may, by diffusion, be eventually transported to a resonance island where
it experiences a large jump in amplitude. Exactly on resonance,
e.g. at $\nu=0.25$, a particle may, after a long time, diffuse on to
a star shaped curve  which subsequently leads to a very large amplitude 
excursion. Particles are most likely to leave a given level curve in the 
vicinity  of the turning points since most time is spent in this neighbouhood.
The projection of diffusive motion along the gradient to the level curves
then transports particles to larger amplitudes.

To analyze the diffusive motion we observe that without noise 
and even after averaging the linear invariant $J$ is not conserved near 
resonance. Instead, after averaging over the fast varying phases,
a time-independent  Hamiltonian is obtained which describes 
motion close to the $2k$th integer resonance (tune $\nu_{2k}$ = 
Integer$/2k$),
\begin{equation}
H \! = \! \delta J + A[F_0+
 2\sum_{m=1}^{\infty} \!\! F_{mk}\cos 2mk\phi]\; .
\end{equation}
where $\delta = \nu^0 - \nu_{2k}$, the difference from the resonance tune,
$\phi = \psi - \nu_{2k}\theta$, the slowly varying phase, and 
$A=C/(2\pi)$. $J$ oscillates 
periodically betweeen two limits $J_{min}$ and $J_{max}$ which
are determined by the transverse energy $E=H$. 
Dropping all $F_{mk},\; m > 1$, the
period on a curve labelled by $E$ is 
$T_E = (1/2\pi k)\int_{J_{min}(E)}^{J_{max}(E)} dJ/
\sqrt{4A^2F_k^2 - (E-\delta J - AF_0)^2}$. In contrast to the 
betatron tune $\nu$, this tune $\nu_E=1/T_E$ is very small, typically of the 
order of 0.001 for $k=2$ and $\delta \sim 10^{-4}$. $\nu_E$ 
increases with the resonance order $2k$. The tune fluctuations 
$\Delta\nu_r$ cause the resonant amplitude and the island widths to also
fluctuate and the energy to diffuse. 
The rate of change of $H$ is found to be $ dH/d\theta = -(dJ/d\theta)_u 
\Delta\nu_r$, the subscript $u$ denotes the unperturbed rate of change. 
After integrating over a turn, the total change 
in the Hamiltonian at turn $N$
is $H(N)-H(0) = -\sum_{m=1}^N [J(m)-J(m-1)]\Delta\nu_r(m)$. Expanding 
$J$ in a Fourier series: $J(m) = \sum_{j=0}^{\infty}B_j \cos (2\pi\nu_E jm +
\theta_j)$ and using the stationarity of 
$\langle \Delta\nu_r(l)\Delta\nu_r(l+n)\rangle$, we obtain for
the diffusion of the energy
\begin{eqnarray}
D_{\nu}(E) \!\!\!& = & \!\!\!\! \frac{(\Delta\nu_r)^2}{1-1/(2\tau_c)} \!\! 
\sum_{j=1}^{\infty} 
\frac{B_j^2 (1 - \cos 2\pi j\nu_E)
{\rm sinh}\Theta}{{\rm cosh}\Theta -\cos 2\pi j\nu_E}
\end{eqnarray}
The Fourier amplitudes $B_j$ grow as the tune approaches
the resonant value from below. The diffusion at a given tune increases 
smoothly moving out from the 
origin, jumps when the particle is on the largest of the resonant islands,
decreases to zero at the stable fixed point, increases back to the value
on the largest island and stays nearly constant thereafter.
The noise frequencies which 
contribute to the diffusion in energy are the harmonics of the
low frequency $\nu_E f_{rev}$.
{\em The topology of the phase space orbits and the fact that noise of 
comparatively low 
frequencies has the dominant contribution to the diffusion in energy 
explains the large growth in emittance due to noise 
in the neighbourhood of a resonance.}

For the 1D beam-beam interaction, diffusion is significantly enhanced
only close to low order resonances, but for the 2D interaction 
even relatively high order resonances e.g. 14th order can lead 
to large emittance growth, as was observed in \cite{Brink}. A detailed study of
the 2D case will appear separately \cite{SenEll}.

Next we consider fluctuations of the offset between the beams at the IP. 
This has two effects. The position of the maximum of the beam-beam force
fluctuates so more particles in the proton beam will be subjected to a larger
force. It also destroys the symmetry of the beam-beam force and can excite odd 
order resonances. This had been studied earlier by
Stupakov \cite{Stup} assuming the potential for a flat beam. Our results
below assume the general form of the potential given in Eq. (\ref{poten}).
We assume that the offset fluctuation $d_r(m)$ at turn $m$ is small and write
it as $d_r(m) = \Delta d_r \chi(m) \sigma_{op} $
where $\Delta d_r$ is the dimensionless amplitude 
of the offset and $\chi(m)$ is a random variable of zero mean and unit 
variance. Calculation of the 
diffusion coefficient far from resonances yields
\begin{eqnarray}
D_{off}(J) &=& \frac{1}{2} (C \sigma_{op} \Delta d_r )^2 \sum_{k=0}^{\infty} 
 (2k+1)^2 G_k^2(a)
 \nonumber \\
& &  \times \sum_{n=-\infty}^{\infty}\!\!\! K_{off}(n) \cos 2\pi(2k+1)\nu n
\end{eqnarray}
The correlation function is $K_{off} = \langle \chi(l)\chi(n+l)\rangle$. 
$G_k$, the Fourier coefficients of the beam-beam force, are given by
$G_k = \sqrt{a}[F_{k+1}'+F_k']/\sigma_{op}
       + [(k+1)F_{k+1}-kF_k]/(\sqrt{a}\sigma_{op}).$
Notice here that the odd harmonics of the betatron tune contribute to the
diffusion in action.

We now discuss the diffusion due to beam size fluctuation of the opposing beam.
The location of the maximum of the beam-beam force is $\propto \sigma_{op}$ 
while the maximum itself is $\propto \sigma_{op}^{-1}$. 
Consequently when $\sigma_{op}$ fluctuates, protons in a larger range of
amplitudes will be subject to the maximum of the force - as with offset
fluctuations.
A study of this for the flat beam potential was reported recently in 
\cite{Koga}. We find that the diffusion coefficient for the general beam-beam 
potential is
\begin{equation}
D_{\sigma_{op}}(J) \!\!= \!\! 32 (C\Delta\sigma_r\ a)^2 
\sum_{k=1}^{\infty}[k F_k']^2 \!\!
\sum_{n=-\infty}^{\infty} \!\! K_{\sigma_{op}} \cos 4\pi k\nu n
\end{equation}
The size fluctuation at turn $n$ is
$\Delta\sigma_r \eta(n) \sigma_{op}$ where $\Delta\sigma_r$ is dimensionless,
$\eta(n)$ is a random variable of mean zero and unit variance, and 
$K_{\sigma_{op}}(n) = \langle \eta(l)\eta(n+l)\rangle$.

\begin{figure}
\centerline{\psfig{figure=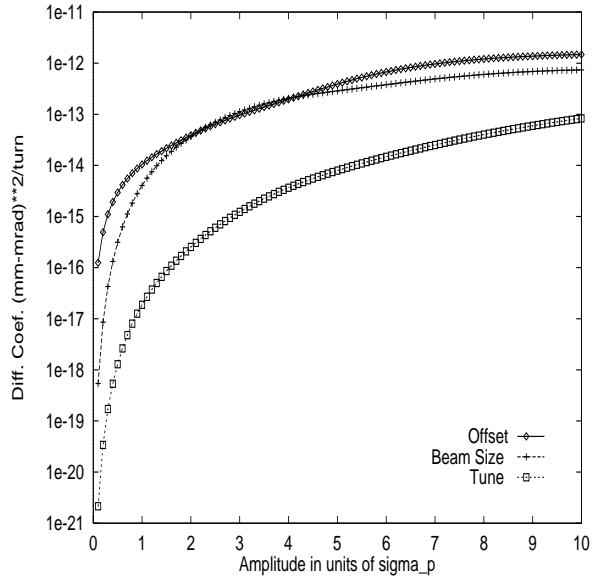,height=3.in,width=3.15in,angle=-90}}
\caption{Comparison of the diffusion coefficients in $J$ due to tune 
fluctuations, offset fluctuations and beam size fluctuations given by 
Equations (5), (9) and (10) respectively.
Parameter values: $\nu^0$=0.291, $\tau_c=10.0$, $\Delta\nu_r=10^{-4}$, 
$\Delta d_r = 0.01 = \Delta \sigma_r$. }
\end{figure}

Figure 3 compares the analytical diffusion coefficients from the three
fluctuating phenomena considered here. The parameter values are shown in the 
caption. We find that the diffusion due to offset fluctuations is 
largest for amplitudes less than 2$\sigma_p$.
At greater amplitudes, diffusion due to beam size 
and offset fluctuations, both of which directly affect the amplitude of the 
beam-beam kick, are of the same order of magnitude \cite{Kog2}.
Diffusion due to tune fluctuations is the smallest at all amplitudes
because it affects only the phase at which the particle is kicked.
Nevertheless, the sources of tune fluctuations are difficult to eliminate 
and more numerous than for the other fluctuations.

To summarise, we have reported three main results in this letter. Far from 
low order resonances, high frequency tune fluctuations cause larger
growth of particle amplitudes than low frequency fluctuations. These high
frequency fluctuations can cause the emittance to nearly double over the
storage time of a day. Near resonances, low frequency fluctuations are
resonant with the motion of the linear invariant and these lead to the largest
diffusion in the energy which subsequently leads to significant emittance 
growth. Comparing different fluctuations in the off-resonance case, we have 
found that for reasonable 
values of the fluctuating amplitudes, offset fluctuations at the interaction 
points cause the largest diffusion at small amplitudes while at large 
amplitudes, fluctuations in the size of the opposing beam 
have a comparable effect as the offset fluctuations.

We thank A. Bazzani, H. Mais and F. Willeke for very fruitful discussions and
support.


\begin{thebibliography}{99}
\bibitem{Kog2} The reasons advanced in ref.\cite{Koga} for the dominance of
the beam size fluctuations are not generally valid since these are the
fluctuations of the {\em opposing} beam.
\end{thebibliography}
\end{document}